\documentclass[a4paper,fleqn,usenatbib]{mnras}
\usepackage[T1]{fontenc}
\usepackage{ae,aecompl}
\usepackage{graphicx}
\usepackage{amsmath}	
\usepackage{amssymb}
\usepackage{color}
\usepackage{mathptmx}
\usepackage{epsfig}
\usepackage{wasysym}
\usepackage{xfrac}
\usepackage{pifont}

\title[Progenitor bias]{Identifying the progenitors of present-day early-type galaxies in observational surveys: correcting `progenitor bias' using the Horizon-AGN simulation}

\author[G. Martin et al.]{
G. Martin,$^{1}$\thanks{E-mail: g.martin4@herts.ac.uk}
S. Kaviraj,$^{1}$
J. E. G. Devriendt,$^{2}$
Y. Dubois,$^{3}$
C. Pichon,$^{3,4}$
and
C. Laigle$^{2}$
\\
$^{1}$Centre for Astrophysics Research, School of Physics, Astronomy and Mathematics, University of Hertfordshire, College Lane, Hatfield AL10 9AB, UK\\
$^{2}$Dept of Physics, University of Oxford, Keble Road, Oxford OX1 3RH UK\\
$^{3}$Institut d'Astrophysique de Paris, Sorbonne Universit\'{e}s, UMPC Univ Paris 06 et CNRS, UMP 7095, 98 bis bd Arago, 75014 Paris, France\\
$^{4}$Korea Institute of Advanced Studies (KIAS) 85 Hoegiro, Dongdaemun-gu, Seoul, 02455, Republic of Korea
}

\pubyear{2017}

\begin{document}
\label{firstpage}
\pagerange{\pageref{firstpage}--\pageref{lastpage}}
\maketitle



\begin{abstract} 
As endpoints of the hierarchical mass-assembly process, the stellar populations of local early-type galaxies encode the assembly history of galaxies over cosmic time. We use Horizon-AGN, a cosmological hydrodynamical simulation, to study the merger histories of local early-type galaxies and track how the morphological mix of their progenitors evolves over time. We provide a framework for alleviating `progenitor bias' -- the bias that occurs if one uses only early-type galaxies to study the progenitor population. Early-types attain their final morphology at relatively early epochs -- by $z\sim1$, around 60 per cent of today's early-types have had their last significant merger. At all redshifts, the majority of mergers have one late-type progenitor, with late-late mergers dominating at $z>1.5$ and early-early mergers becoming significant only at $z<0.5$. Progenitor bias is severe at all but the lowest redshifts -- e.g. at $z\sim0.6$, less than 50 per cent of the stellar mass in today's early-types is actually in progenitors with early-type morphology, while, at $z\sim2$, studying only early-types misses almost all (80 per cent) of the stellar mass that eventually ends up in local early-type systems. At high redshift, almost all massive late-type galaxies, regardless of their local environment or star-formation rate, are progenitors of local early-type galaxies, as are lower-mass (M$_\star$ $<$ 10$^{10.5}$~M$_{\odot}$) late types as long as they reside in high density environments. In this new era of large observational surveys (e.g. LSST, JWST), this study provides a framework for studying how today's early-type galaxies have been built up over cosmic time.
\end{abstract}




\section{Introduction}
In the standard $\Lambda$CDM paradigm, galaxy formation proceeds hierarchically. Dark matter halos, which arise as a result of primordial fluctuations in the initial matter density field \citep{Starobinsky1982,Guth1982,Hawking1982}, merge to form progressively more massive haloes over cosmic time \citep[e.g.][]{Blumenthal1984,Kauffmann1993,Somerville1999}. Cold gas condenses into these halos where it forms rotationally-supported discs \citep{Franx1996}. The rate of star-formation is determined by the local density of this cold gas \citep{Kennicutt1998}, with feedback from supernovae \citep{Scannapieco2008} and active galactic nuclei (AGN) \citep[e.g.][]{Silk1998,Kaviraj2016} regulating the process of stellar mass growth. A consequence of this paradigm is that the stellar mass of an individual galaxy is assembled through a combination of in-situ star-formation i.e. by gas turning into stars within a galaxy's own halo, and ex-situ star-formation i.e. stars formed in another halo which have become members of the halo in question as a result of merging \citep[e.g.][]{Kauffmann1993}. As `end-points' of this hierarchical assembly process, local `early-type' (i.e. elliptical and S0) galaxies are a particularly significant class of objects \citep[e.g.][]{Kaviraj2007}. These galaxies dominate the stellar mass density in today's Universe \citep[e.g.][]{Bernardi2003} and thus encode, in their stellar populations, the signatures of galaxy mass assembly over cosmic time \citep[e.g.][]{Worthey1994,Dokkum2001}. Studying these galaxies offers unique insights into the build up of the observable Universe \citep[e.g.][]{Barrientos2003,Longhetti2005,McDermid2015} and significant effort in the literature has, therefore, been dedicated to understanding these systems. 

Galaxies in the early universe form with disc-like (late-type) morphologies and, through interactions and secular processes, acquire more spheroidal (early-type) morphologies over time \citep[e.g.][]{Franx1996,Dokkum2001,Buitrago2014,Conselice2014,Kaviraj2014}. Consequently, at progressively higher redshift, the progenitors of today's early-types become increasingly dominated by late-type systems. Understanding the formation and evolution of today's early-types therefore requires us to consider their entire progenitor population, especially their late-type progenitors at earlier epochs. If, as is often assumed, early-type galaxies do not revert back to late-types \citep[e.g.][]{Hau2008,Fang2012}\footnote{\citet{Dubois2016} have shown that transformation from early-type to late-type morphology is prevented by AGN feedback.}, techniques that identify \textit{late-type} galaxies that are progenitors of today's early-types in observational surveys are essential. This becomes particularly important at high redshift, e.g. $z>2$, the redshift regime at which the early-type population is rapidly assembled \citep[e.g.][]{Conselice2014}, and which can be routinely accessed by forthcoming facilities such as JWST \citep{Gardner2006}, EUCLID \citep{Laureijs2011}, etc. 

Past observational studies \citep[e.g.][]{Gladders1998,Stanford1998} have attempted to trace the assembly of present-day early-types by focussing only on the population of early-type galaxies at high-redshift. Since the stellar populations of present-day early-types are largely in place in the early Universe \citep[e.g.][]{Trager2000,Thomas2005,Kaviraj2011,Kaviraj2014b}, this is a reasonable approximation at low and intermediate redshift ($z\sim1$), assuming that galaxies, once they achieve early-type morphology, cannot revert back to being late-type systems. However, as noted above, at earlier epochs, an increasing proportion of the mass in present-day early-types is contained in late-type progenitors. Thus, considering only early-type galaxies introduces a bias (the so-called `progenitor bias') in any study of their evolution, which becomes progressively more severe with increasing redshift. 

In a similar vein, other observational work \citep[e.g.][]{Bell2004} has often used the optical red sequence \citep{Faber2007,Barro2013} as a proxy for the population of progenitors of early-type galaxies. However, since the stellar populations in late-type galaxies tend to be younger (and therefore bluer), such a colour cut misses the majority of the late-type galaxies that are progenitors of local early-types. In addition, since a wide variety of star-formation histories are observed in early-type galaxies themselves, particularly at high redshift \citep{Kaviraj2013,Fitzpatrick2015,Lofthouse2017}, a large fraction of blue galaxies that already have early-type morphology will also be missed \citep{Shankar2015}, if such a colour cut is employed. While the red sequence traces the progenitors of early-type galaxies well at the highest end of the luminosity function \citep{Kaviraj2009}, it becomes less reliable at low masses (where galaxies are bluer) and fails to identify early and late-type progenitors (of all stellar masses) that lie blueward of the red sequence. In a general sense, therefore, a simple colour cut is not a reliable approximation for the progenitor population of today's early-type galaxies. And, in a similar vein to using an early-type selection to identify the progenitor population, the red-sequence approximation becomes progressively less effective with increasing redshift. 

Progenitor bias is difficult to overcome observationally, since individual galaxies cannot be observed as they evolve and, therefore, in any given survey, it is difficult to directly identify objects that will end up in early-type galaxies at $z\sim 0$. Nevertheless, observational methods have been applied to reduce or mitigate this issue. For example, a widely-used technique, proposed by \citet{Dokkum2010} assumes that cumulative co-moving number density is conserved as the galaxy mass function evolves with time (i.e. as a galaxy evolves, the number of galaxies more massive than it remains constant), so that a given galaxy maintains the same rank. However, while this method is able to account for the mass evolution of the galaxy population, the assumptions made may be too simplistic. For example, it ignores the fact that a galaxy's rank may change over time, resulting in the evolution of its co-moving number density. This may occur if, for instance, galaxies are removed from the population as a result of mergers \citep{Ownsworth2014}, or as a result of the spread in specific star-formation rate (sSFR) and its dependence on mass \citep{Leja2013,Shankar2015}.

Improvements on the method of \citet{Dokkum2010} which include prescriptions for the number density evolution of galaxies have been proposed. For example, \citet{Behroozi2013} uses abundance matching in order to match observed galaxies to corresponding dark matter halos in $\Lambda$CDM simulations, allowing the median and dispersion of the cumulative co-moving number density tracks to be quantified for the progenitors of galaxies of a given mass. Work by \citet{Torrey2015,Torrey2017} introduces an analytic framework, which accounts for the effects of the merger rate (`coagulation') and stochasticity of galaxy growth rates (`scatter rate'), and describes the median \citep{Torrey2015} and intrinsic scatter \citep{Torrey2017} of the evolution of the galaxy population in co-moving density space. \citet{Wellons2017} present a probabilistic method based on this framework, which they show is able to more effectively predict progenitor properties than the methods of \citet{Dokkum2010} and \citet{Behroozi2013}.

While this group of methods is widely applicable to any given galaxy property \citep[e.g.][]{Torrey2015,Clauwens2016}, they may lose predictive power in cases where there is no expectation that the rank of the property of interest will be conserved (e.g. morphology). In such cases, leveraging the constraining power of additional galaxy properties becomes essential. Additionally, all of these methods retain a weakness of the \citet{Dokkum2010} method, in that they still assume that early-type progenitors follow the same distribution of properties as the general population at a given redshift.

An appealing alternative is to employ a simulation that reproduces the properties of galaxies over cosmic time. Since the identities of the progenitors of local early-type galaxies are precisely known in the model, they can be used to calculate the probability that a galaxy is a progenitor, as a function of its observable properties (e.g. redshift,  stellar mass, star-formation rate (SFR), local environment). \citet{Kaviraj2009} have previously addressed the problem of progenitor bias using the GalICS semi-analytical model \citep{Hatton2003}. However, while the semi-analytical approach has successfully reproduced the phenomenology of many aspects of the galaxy formation process \citep[e.g.][]{Somerville1999,Cole2000,Benson2003,Bower2006,Croton2006}, the recent advent of hydrodynamical simulations in cosmological volumes provides a more accurate route to addressing the problem. Unlike their semi-analytical counterparts, hydrodynamical models resolve the gas and baryonic content of galaxies, typically on kpc scales, \citep[e.g.][]{Devriendt2010,Dubois2014,Vogelsberger2014,Schaye2015,Khandai2015,Taylor2016,Kaviraj2016}. This enables them to more accurately model a greater range of physical processes, without the need for semi-analytical recipes \citep{Schaye2015}, although some processes, such as AGN feedback, that cannot be resolved, must still be described using sub-grid models \citep[e.g.][]{Katz1992,Booth2009,Kimm2015}. Such hydrodynamical simulations typically rely on a smaller number of free parameters and offer a more realistic treatment of the physical processes involved in the formation and evolution of galaxies, yielding better agreement with the observed Universe without the need for tuning. 

\begin{figure}
  \centering
  \includegraphics[width=0.5\textwidth]{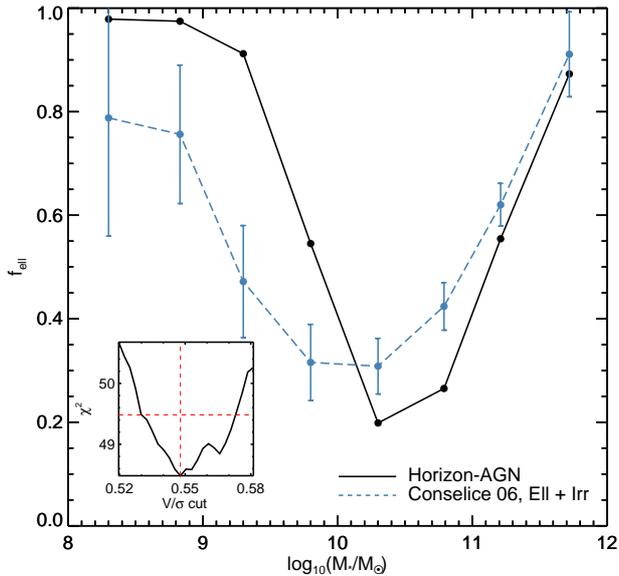}
  \caption{The predicted early-type fraction in Horizon-AGN, for the $\sfrac{V}{\sigma}$ value (0.55$\pm^{0.03}_{0.02}$) that best reproduces the observed early-type fractions in the local Universe \citep{Conselice2006}. The inset shows the value of $\chi^{2}$ between the predicted and observed early-type fractions for different $\sfrac{V}{\sigma}$ thresholds (the red dashed lines show the minimum $\chi^{2}$ and $\Delta\chi^{2} = 1$).}
  \label{vscompare}
\end{figure}

In this study, we use the Horizon-AGN\footnote{\url{http://www.horizon-simulation.org}}  cosmological simulation \citep{Dubois2014,Kaviraj2016} to (1) quantify the evolution of the progenitor population of today's early-type galaxies and (2) provide a route for identifying late-type galaxies that are progenitors of present-day early-types in observational surveys, by estimating the probability of a given late-type to be the progenitor of a local early-type system, as a function of measurable observables like redshift, stellar mass, star-formation rate and local density. 

The structure of this paper is as follows. In Section \ref{sec:Dataset}, we describe the main characteristics of the simulation, describe our simulated galaxy sample and define how observables are measured. In Section \ref{sec:Redshiftevolution}, we probe the redshift evolution of the progenitors of present-day early-type galaxies. In Section \ref{sec:Progenitorfractions}, we present probabilistic prescriptions to identify late-type galaxies that are progenitors of local early-types as a function of redshift, stellar mass, environment and star-formation rate. We summarize our findings in Section \ref{sec:Summary}. 

\begin{figure*}
\begin{minipage}{172mm}
  \centering
  \includegraphics[width=1.0\textwidth]{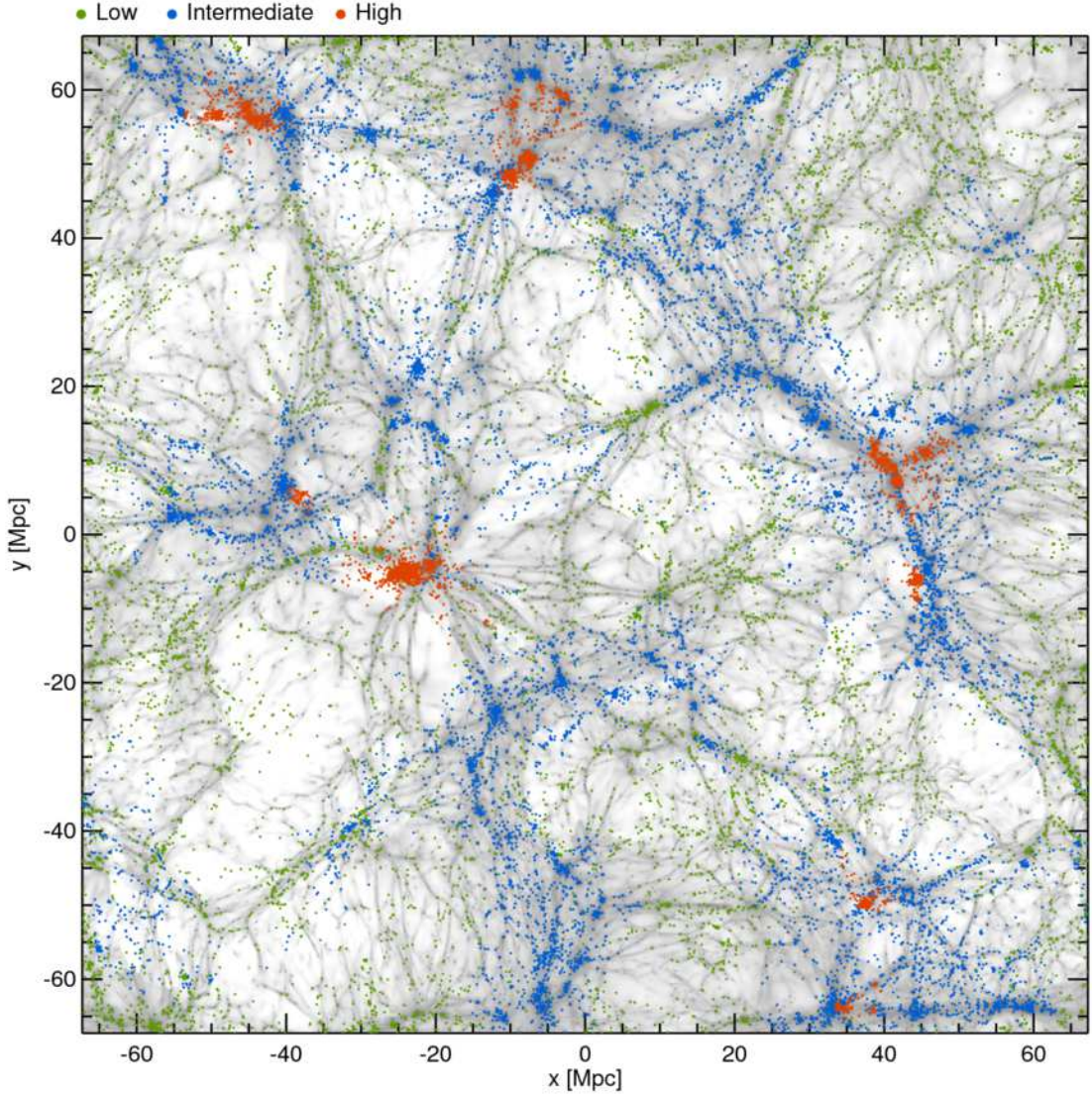}
  \caption{Greyscale map showing the projected gas density in a $\sim$ 20~Mpc slice through the simulation volume at the final time output ($z = 0.06$). Green, blue and red points correspond to the positions of galaxies ($\mathrm{M_\star} > 10^{9.5}~\mathrm{M}_{\astrosun}$) in low, intermediate and high density environments respectively (see text in Section \ref{sec:densityestimation} for more details). Axes are in units of proper Mpc.}
  \label{gasmap}
\end{minipage}
\end{figure*}

\section{The simulation}
\label{sec:Dataset}
We begin by briefly describing the Horizon-AGN simulation, the extraction of galaxies and merger trees and the prediction of observable quantities. 

\subsection{Horizon-AGN}
Horizon-AGN is a cosmological hydrodynamical simulation \citep{Dubois2014} that employs the adaptive mesh refinement Eulerian hydrodynamics code, RAMSES \citep{Teyssier2002}. The size of the simulation box is 100~$h^{-1}\, \rm coMpc$, which contains $1024^3$ dark matter particles and uses initial conditions from a \emph{WMAP7} $\Lambda$CDM cosmology \citep{Komatsu2011}. The simulation has a dark matter mass resolution of $8\times10^7$~M$_{\odot}$, a stellar-mass resolution of $2\times10^6$~M$_{\odot}$ and a spatial resolution of $\sim$1~kpc. We direct readers to \citet{Kaviraj2016} for details of the recipes (e.g. star-formation and stellar and AGN feedback) employed to model the baryonic evolution of galaxies. Briefly, star-formation follows a standard Schmidt-Kennicutt law \citep{Kennicutt1998}, with the model implementing continuous stellar feedback, that includes momentum, mechanical energy and metals from stellar winds, Type II SNe and Type Ia SNe. Black-hole (BH) feedback on ambient gas operates via two separate channels, depending on the gas accretion rate. For Eddington ratios $>0.01$ (high accretion rates), 1.5 per cent of the accretion energy is injected as thermal energy (a quasar-like feedback mode), whilst for Eddington ratios $<0.01$ (low accretion rates), bipolar jets are employed with a 10 per cent efficiency. The parameters are chosen to produce agreement with the local cosmic black-hole mass density, and the M$_{\rm BH}$ -- M$_\star$ and M$_{\rm BH}$ -- $\sigma_\star$ relations \citep{Dubois2012}. Apart from choosing the BH-feedback parameters to match the M$_{\rm BH}$ -- M$_\star$ and M$_{\rm BH}$ -- $\sigma_\star$ relations at $z=0$, Horizon-AGN is not otherwise tuned to reproduce the bulk properties of galaxies at $z\sim0$. As described in \citet{Kaviraj2016}, the simulation reproduces key quantities that trace the aggregate stellar mass growth of galaxies: stellar mass and luminosity functions, rest-frame UV-optical-near infrared colours, the star-formation main sequence and the cosmic star-formation history. 

\subsection{Identifying galaxies and building merger trees}
In order to track their progenitors, we build merger histories for each early-type galaxy in the final snapshot of the simulation ($z=0.06$). In the sections below, we describe the process of galaxy identification, followed by the process of building merger trees. 

\subsubsection{Identifying the galaxy sample}
For each snapshot, we produce a catalogue of galaxies, using the \textsc{AdaptaHOP} structure finder \citep{Aubert2004,Tweed2009}, operating directly on the star particles. Galactic structures are selected using a local threshold of 178 times the average matter density, where the local density of individual particles is calculated using the 20 nearest neighbours. Only galactic structures with more than 100 star particles are considered, corresponding to a minimum galaxy stellar mass of M$_{\star}\sim2\times$10$^{8}$~M$_{\astrosun}$, which is a consequence of a minimum star particle mass of $m_{\star} \sim 2\times10^{6}$~M$_{\astrosun}$. We identify an average of $\sim$150,000 galaxies above the 100 particle threshold in each snapshot. We restrict our study to galaxies with stellar mass M$_{\star}>10^{9.5}$~M$_{\astrosun}$, for reasons outlined in Section \ref{sec:Mergerhistories}.

\subsubsection{Producing merger histories}
\label{sec:Mergerhistories}
Using the catalogue of galaxies identified by \textsc{AdaptaHOP}, we extract merger histories for each early-type galaxy at the final snapshot. We produce merger trees using 91 snapshots in the range $z \in [0.06,7.04]$, with an average time-step of $\sim$130~Myr. Merger trees are produced for each early-type, by identifying their progenitors at each snapshot, using the method described in \citet{Tweed2009}. Since our threshold for identifying structures is 100 star particles, only mergers where the satellite galaxy has M$_\star\gtrsim 2\times10^{8}$~M$_{\astrosun}$ are considered, regardless of mass ratio. Given that our sample excludes galaxies less massive than M$_{\star} \sim 10^{9.5}$~M$_{\astrosun}$, and the minimum galaxy mass identified is M$_{\star} \sim 2\times10^{8}$~M$_{\astrosun}$, our sample is complete for mergers down to a mass ratio of at least 1:15. 

\subsection{Prediction of observables}
We produce observables that can be used in conjunction with contemporary and future observational datasets. These are stellar mass (derived using the total mass of the star particles in a galaxy), star-formation rate, local number density and stellar kinematics (which we use as a proxy for morphology). The following sections describe how we derive each of these measures. 

\subsubsection{Morphology}
\label{sec:morphology}
The morphology of each model galaxy in our analysis is inferred using its stellar kinematics. Morphology is defined using $\sfrac{V}{\sigma}$, which is the ratio of the mean rotational velocity ($V$) to the mean velocity dispersion ($\sigma$), both measured using the entire star particle distribution. Higher values of this ratio correspond to more late-type (disc-like) morphologies. $\sfrac{V}{\sigma}$ is calculated by first rotating the coordinate system so that the $z$-axis is oriented along the angular momentum vector of the galaxy. Rotational velocity is defined as the mean tangential velocity component in cylindrical co-ordinates, $V_{\theta}$, and velocity dispersion is computed using the standard deviations of the radial, tangential and vertical star particle velocities, $\sigma_{r}, \sigma_{\theta}$ and $\sigma_{z}$, summed in quadrature. $\sfrac{V}{\sigma}$ is then given by

\begin{equation}
\sfrac{V}{\sigma} = \frac{\sqrt{3} \bar{V}_{\theta}}{\sqrt{\sigma^{2}_{r}+\sigma^{2}_{\theta}+  \sigma^{2}_{z}}}.
\end{equation}

To separate galaxies morphologically into early and late types using $\sfrac{V}{\sigma}$, we consider a range of values for $\sfrac{V}{\sigma}$ and compare how the resulting predicted early-type fractions compare to their observed counterparts at low redshift. The value of $\sfrac{V}{\sigma}$ which produces the best agreement with the observational data is then selected as the threshold value that separates early-types and late-types in the model. Figure \ref{vscompare} shows the predicted early-type fractions in Horizon-AGN for this best-fitting $\sfrac{V}{\sigma}$ value, 0.55$\pm^{0.03}_{0.02}$, compared with early-type fractions derived from observations \citep{Conselice2006}. The largest discrepancy between the observations and the simulation occurs at the low mass end, and is likely a result of insufficient mass resolution \citep{Dubois2016}. Nevertheless, over the mass range considered in this study ($\mathrm{M_\star} > 10^{9.5}~\mathrm{M}_{\astrosun}$), the early-type fractions predicted by Horizon-AGN, for our $\sfrac{V}{\sigma}$ threshold of 0.55, are in reasonable agreement with the observations. We note that the progenitor fractions presented in Section \ref{sec:Progenitorfractions} are resistant even to relatively large changes in our $\sfrac{V}{\sigma}$ threshold. Varying the $\sfrac{V}{\sigma}$ threshold by as much as 50 per cent introduces only a $\pm 0.05$ variation in our calculated progenitor fractions. 

We note that the minimum refinement of the AMR grid is increased at $z=$4 ,1.5 and 0.25, in order to keep the minimum physical cell size approximately constant \citep{Dubois2014,Perani2016}. While it is possible that this refinement may result in the production of sudden instabilities in previously stable discs that increase galaxy velocity dispersions, the smooth nature of the $\sfrac{\mathrm{V}}{\sigma}$ evolution of the galaxy population \citep{Dubois2014}, indicates that this is not a significant effect.  

\begin{figure*}
  \centering
  \includegraphics[width=1.0\textwidth]{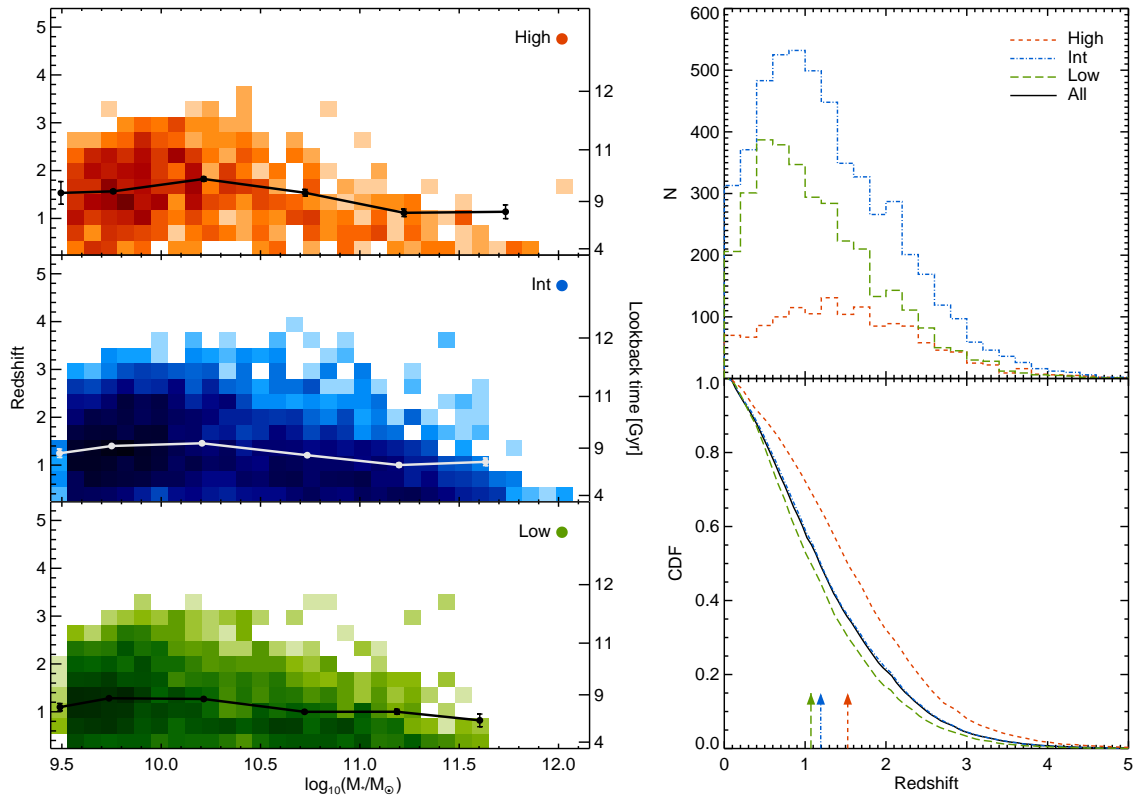}
  \caption{Left: Density plots showing the redshift at which local early-type galaxies in different environments had their last significant merger (i.e. a merger with a mass ratio greater than 1:10). Green, blue and red  colours correspond to low, intermediate and high density environments. Over-plotted are lines showing the mean redshift in each bin and its error. Right: Histograms (top) and associated cumulative distribution functions (bottom) of last merger redshifts. The arrows in the bottom panel indicate the redshift where the cumulative distribution functions reach values of 0.5 (i.e the median value) for each environment.}
  \label{mergerscatter}
\end{figure*}

\subsubsection{Local environment}
\label{sec:densityestimation}
We compute the local environment of each galaxy using an estimate of the 3-D local number density. For each galaxy, we estimate the surrounding number density of galaxies above our mass cut ($10^{9.5}$~M$_{\astrosun}$). This is achieved using the adaptive kernel density estimation method of \citet{Breiman1977} which utilises a finite-support Epanechnikov kernel (instead of the more typically-used infinite-support Gaussian kernel) and using the density estimator itself as a pilot to steer the local kernel size \citep{Wilkinson1995,Ferdosi2011}. Galaxies at our final snapshot ($z=0.06$) are classified into `high', `intermediate' and `low' density environments, by comparing the density percentile they occupy to the corresponding percentiles that observed galaxies inhabiting cluster, group and field environments occupy in the low-redshift universe.

Observations indicate that around 10 per cent of galaxies occupy rich clusters, while around 50 per cent of galaxies occupy groups and poor clusters \citep[e.g][]{Bahcall1996,Dekel1999,Tempel2012}, with the remaining 40 per cent of galaxies occupying the field. We use these values as a guide to make density cuts, in order to separate our simulated galaxies into our three density bins (high, intermediate and low), that are roughly analogous to cluster, group and field environments. To allocate our simulated galaxies into these density bins, we first rank the objects by density. Then, the 10 per cent of galaxies with the highest local density (i.e. the 90$^{\mathrm{th}}-100^{\mathrm{th}}$ percentile range) are classified as occupying high density environments, galaxies in the 40$^{\mathrm{th}}-90^{\mathrm{th}}$ range are classified as occupying intermediate density and the remaining galaxies are classified as occupying low-density environments. Figure \ref{gasmap} shows the environment classifications of galaxies with $\mathrm{M_\star} > 10^{9.5}~\mathrm{M}_{\astrosun}$ superimposed over a map of the gas density. Not unexpectedly, galaxies classified as being in high density environments lie at the nodes of the cosmic web, while intermediate density galaxies lie largely along filaments, with low-density galaxies typically being found in voids. Regions of high gas density also host a high number density of galaxies which, as we show later in Section \ref{sec:Redshiftevolution}, leads to more rapid morphological transformation. More massive galaxies (those above our detection threshold of $2\times10^{8}~\mathrm{M}_{\astrosun}$) first appear in the simulation around nodes, owing to a greater abundance of gas from which to accrete, and also undergo a higher incidence of interactions or mergers. These galaxies are thus likely to experience earlier and accelerated evolution compared to galaxies in less dense environments.

In all of our analysis below, we always cast the local environment in terms of the density percentile of individual galaxies. This is because observers (and theorists) inevitably use different metrics for measuring local density. However, while the absolute values of density depends on the actual metric being used, the density percentile that a galaxy occupies is likely to be roughly independent of the actual estimation method (as we demonstrate later in Section \ref{sec:Progenitorfractions}). 

\subsubsection{Star-formation rate}
Star-formation rates are calculated by computing the change in stellar mass, $m_{\star}$, of the galaxy in question between two snapshots, dividing by the time difference between those snapshots, and subtracting the mass of stars formed \emph{ex-situ} that has merged with the galaxy between the two snapshots, $m_{\star,merged}$:

\begin{equation}
  \mathrm{SFR} = \frac{m_{\star,t=t_{i}}-m_{\star,t=t_{i-1}}-m_{\star,merged}}{\Delta t}, 
\end{equation}

where $t_{i}$ is the time at the current snapshot and $t_{i-1}$ is the time at the previous snapshot. $\Delta t$ is equal to $t_{i}-t_{i-1}$ and, as discussed in Section \ref{sec:Mergerhistories}, the average time-step used in the simulation is $\sim$130~Myr. We note that the SFR in not sensitive to the exact value of $\Delta t$. For example, if we double or halve the value $\Delta t$, in either case, our calculated SFRs change by less than 30 per cent.

\begin{figure}
  \centering
  \includegraphics[width=0.5\textwidth]{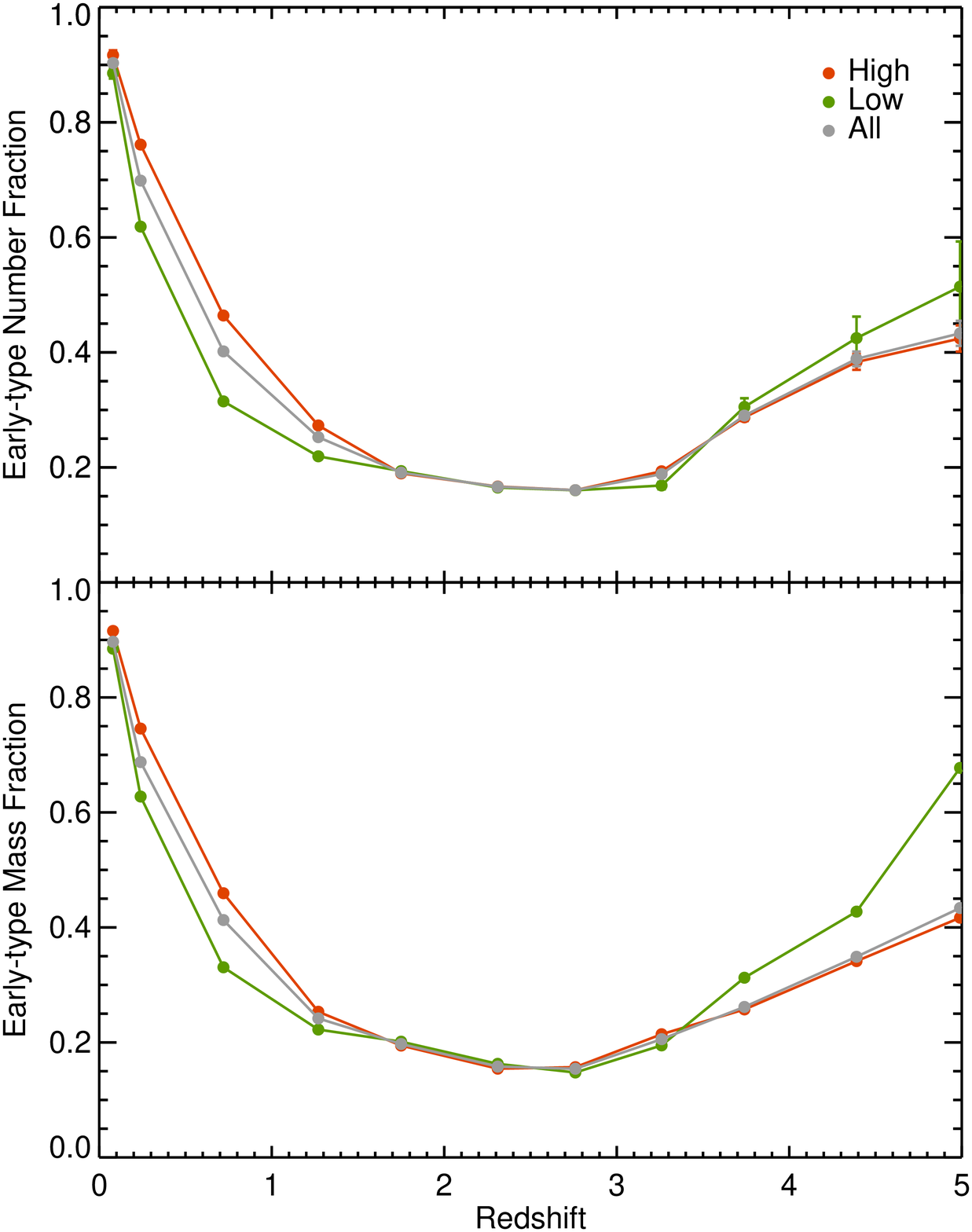}
  \caption{Top: The number fraction of progenitors that already have early-type morphology, split by the local environment of the early-type galaxy that the progenitor ends up in at the present day. Bottom: The mass fraction contained in progenitors that have early-type morphology, split by the environment of the early-type galaxy that the progenitor ends up in at the present day. Poisson error bars are shown (note that most lie within the data points).}
  \label{progenitorenvironment}
\end{figure}

\section{Redshift evolution of the progenitors of local early-type galaxies}
\label{sec:Redshiftevolution}
We begin our analysis by performing a broad exploration of the redshift evolution of the progenitors of local early-type galaxies in the simulation. The left-hand panel of Figure \ref{mergerscatter} presents density plots showing the last-merger redshift of each local early-type galaxy. We define this as a galaxy's last significant merger, i.e. a merger which has a mass ratio greater than 1:10. The right-hand panels of Figure \ref{mergerscatter} show histograms (top-right) and their associated cumulative density functions (bottom-right) as a function of redshift for each environment. The arrows in the bottom-right panel indicate the redshift where the cumulative distribution functions reach values of 0.5 (i.e. the median value) for each environment.

Local early-type galaxies that inhabit denser environments tend to have higher \textit{dynamical ages} i.e. their last significant mergers take place at earlier epochs. While 50 per cent of galaxies in the high density bin have their last significant merger by $z=1.5$, this is only the case for galaxies in the lowest density bin at $z=1.1$ (as indicated by the coloured arrows in the cumulative density function plot (bottom-right panel in  Figure \ref{mergerscatter})). Local early-types in higher-density environments also tend to have higher final masses indicating that the bulk of their evolution takes place earlier and is more rapid, although within a specific environment (high, intermediate or low), the most massive galaxies finish assembling at later epochs \citep[see also][]{DeLucia2007, Dubois2016}.

In Figure \ref{progenitorenvironment}, we quantify the extent of progenitor bias at various redshifts, as a function of galaxy mass and local environment. We show both the fraction of progenitor galaxies that have already acquired early-type morphology (top panel) and the mass fraction in the progenitor population that is contained in progenitors with early-type morphology (bottom panel). We find that, across all environments, only $\sim 50$ per cent of the progenitors have acquired early-type morphology by $z \sim 0.6$. This is also true of the mass fraction contained in these early-type progenitors i.e. at $z \sim 0.6$ only half of the stellar mass that eventually ends up in early-type galaxies today is contained in progenitors that have early type morphology. In other words, looking \textit{only} at early-type systems to trace the evolution of today's early-types would miss half of the progenitor population at $z\sim0.6$. 

Since morphological transformation is more rapid in regions of higher density, progenitor bias is less severe. Thus, 50 per cent of the progenitors of local early-types in high-density environments (i.e. today's clusters) have already acquired early-type morphology by $z=0.7$ (compared to $z=0.6$ across all environments). Note, however, that the bias remains reasonably high regardless of environment. Very similar trends are seen when quantifying progenitor bias as a function of stellar mass (not shown in Figure \ref{progenitorenvironment}), with more massive galaxies (M$_{\star}>10^{11.5}$~M$_{\odot}$) following the same trend as the `high' density environment in Figure \ref{progenitorenvironment}. This is simply because the most massive galaxies occur overwhelmingly in dense environments.

Finally, we note that the early-type fraction appears to decrease (somewhat counter-intuitively) from $z=5$ and begins to increase again around $z=3$. This is not an artefact of our $2\times10^{8}$~M$_{\odot}$ detection threshold, which might cause the most rapidly evolving (and more massive) galaxies to be detected first, potentially biasing our result. If we limit our study to narrow mass bins or follow only the evolution of galaxies that are detected by the structure finder at $z\sim5$, we observe the same non-monotonic evolution. This is partially the result of generally more clumpy star formation and more disturbed morphologies at high redshift \citep[e.g.][]{Ceverino2010}, and is consistent with observational work \citep[e.g.][]{Kassin2012,Kassin2014}, which has shown that star-forming galaxies steadily settle into flat, rotationally supported discs, through the process of `disc settling' at these epochs. This kinematic settling is driven by the fact that many processes which are able to `puff up' (i.e. increase the dispersional motion) of gas in the disc without disturbing stellar orbits significantly, become less frequent and/or less intense with time - examples of this include (minor) mergers and gas accretion episodes \citep[e.g.][]{Covington2010,Bournaud2011,Lofthouse2017b,Martin2017}, strong stellar feedback as a result of the high star-formation rates at high redshift \citep[e.g.][]{Silk2009} and the high gas fractions at early epochs that lead to increased disc instability. In essence, the more gentle evolution that galaxies undergo at later times is thought to allow the gas in the disc to settle into a more ordered state \citep{Kassin2012}. Star formation then proceeds primarily in a planar disc, gradually reducing the mean $\sfrac{V}{\sigma}$ as more stars form. Indeed, for Horizon-AGN galaxies that still host a significant disc at $z=0.06$, we find that old stars (those formed before $z=3$) are more likely to be found in orbits outside of the plane of the disc, symptomatic of the fact that, at these early epochs, gas fractions and merger rates were typically higher on average. Disc settling has also been observed in other simulations \citep[e.g.][]{Kassin2014,Ceverino2017}. In the case of Horizon-AGN, this effect may be compounded slightly by changes to the maximum refinement of the gas grid, which, as mentioned in Section \ref{sec:morphology} is increased towards lower redshifts in order to keep the minimum physical cell size approximately constant. This can have the effect of artificially thickening discs of some galaxies, although we still observe non-monotonic evolution regardless of refinement level.

\begin{figure}
  \centering
  \includegraphics[width=0.5\textwidth]{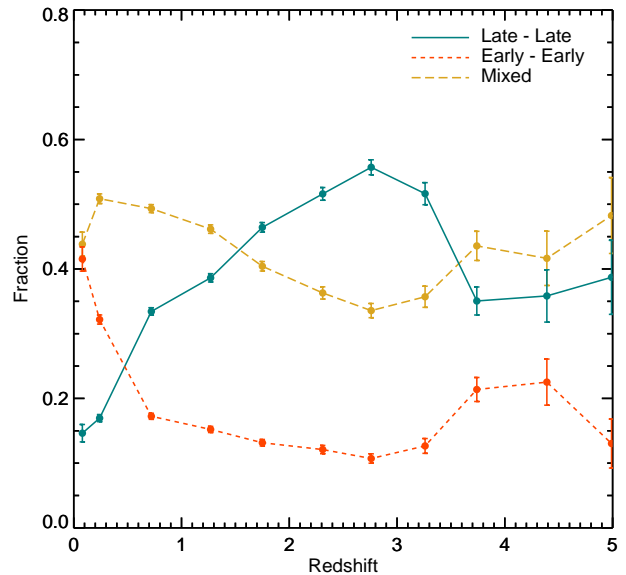}
  \caption{The morphological composition of binary mergers within the progenitor population, that have mass ratios greater than 1:10, as a function of redshift. `Late-late' indicates mergers between two late-type galaxies, `mixed' indicates mergers between one early-type and one late-type galaxy and `early-early' indicates a merger between two early-type galaxies. Poisson errors are shown.}
  \label{morphologyfractions}
\end{figure}

We conclude this section by exploring the morphologies of the progenitors of early-type galaxies that are involved in mergers. We focus only on binary mergers because, although non-binary mergers do occur, they are rare (around two orders of magnitude less frequent than binary mergers). In Figure \ref{morphologyfractions}, we show the fraction of mergers at a given redshift that involve two late-type galaxies (`late-late'), one late-type and one early-type galaxy (`mixed') and two early-type galaxies (`early-early'). Mergers between two late-type galaxies dominate in the early Universe i.e. around the epoch of peak cosmic star-formation and beyond ($z>1.5$). The fraction of mergers involving two early-type galaxies climbs rapidly in the low redshift Universe ($z<0.5$). However, at all redshifts, the majority of mergers involve at least one late-type galaxy. 

\section{Identifying late-type galaxies that are progenitors of present day early-types}
\label{sec:Progenitorfractions}
We proceed by constructing probabilistic prescriptions for identifying late-type progenitors of local early-type galaxies in observational surveys, as a function of quantities that are measurable in today's datasets: redshift, stellar mass, local environment and star-formation rate. As noted in the introduction, the overall aim is to provide a means for correcting progenitor bias in observational studies, by allowing for the inclusion of late-type progenitors of today's early-type systems. We do this by calculating the fraction of late-type galaxies that are progenitors of local early-types, as a function of the measurable quantities mentioned above. This fraction can then be thought of as a probability that a galaxy with the given properties is the progenitor of an early-type galaxy at present day. Observers who wish to include late-type progenitors of early-type galaxies can then use these probabilities to `weight' objects in observational surveys, thus enabling them to reduce progenitor bias by including, in a probabilistic sense, the late-type members of the progenitor population. These probabilistic prescriptions are likely to be particularly useful in the new era of deep-wide surveys (e.g. DES \citep{DES2005}, EUCLID \citep{Laureijs2011}, LSST \citep{Anderson2009}, JWST \citep{Gardner2006} etc.) which will routinely offer large datasets that probe the early Universe, where progenitor bias becomes most severe, and simplifying assumptions, such as using only early-type galaxies or the red sequence to trace the progenitor population, break down. In what follows, we first explore 1-D progenitor probabilities as a function of stellar mass (split by local environment) and then 2-D probabilities as a function of mass and environment and mass and star-formation rate.  

\begin{figure*}
  \centering
  \includegraphics[width=1.0\textwidth]{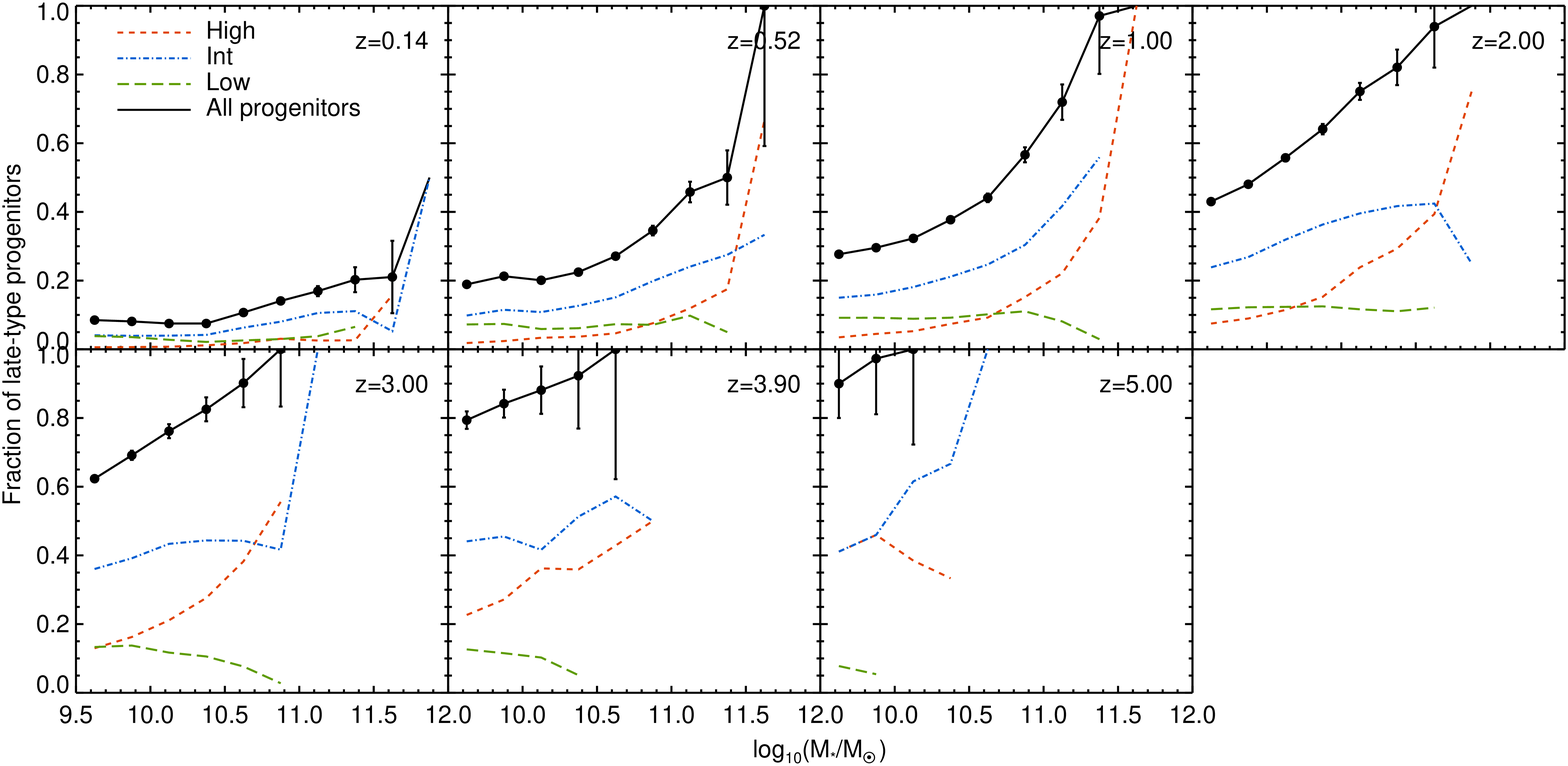}
  \caption{The evolution with redshift of the fraction of late-type galaxies that are progenitors of early-type galaxies at $z \sim 0$. Each panel shows, in black, the fraction of late-type galaxies that are progenitors of local early-types for a given redshift, as a function of the stellar mass of the progenitor. The sample of late-type galaxies is split further into high (red), intermediate (blue) and low (green) density environments. Error bars show Poisson errors. The sample becomes smaller (leading to a corresponding increase in the size of the Poisson error bars) towards higher redshifts, because there are fewer galaxies with stellar masses above M$_{\star}>10^{9.5}$ in the simulation. For clarity, errors bars and points are are not shown where the errors are larger than 0.5.}
\label{progenitorfraction}
\end{figure*}

Since we are interested in probing progenitor probabilities as a function of local environment, we calculate, at each redshift of interest, the 3-D local number density using the method described in Section \ref{sec:densityestimation}. As in Section \ref{sec:Redshiftevolution}, we consider galaxies in the 90$^{\mathrm{th}}-100^{\mathrm{th}}$ percentile range to be inhabiting high density environments, those in the 40$^{\mathrm{th}}-90^{\mathrm{th}}$ percentiles to be inhabiting intermediate-density environments and those in the 0$^{\mathrm{th}}-40^{\mathrm{th}}$ percentile range to be inhabiting low density environments. As noted before, the density percentile in which a galaxy lies (which is driven by its rank in density) is likely to be reasonably resistant to the exact method used for the density estimation. 

To check this, we compare two different density estimation methods. These are the adaptive kernel density estimator used in Section \ref{sec:densityestimation} and the $k^{\mathrm{th}}$ nearest neighbours density estimator, that is commonly used in many observational studies \citep[e.g.][]{Dressler1980,Baldry2006,Ferdosi2011,Shattow2013}. Note that for consistency with the adaptive kernel method, we choose a definition for the $k^{\mathrm{th}}$ nearest neighbour algorithm whereby each galaxy is considered to be its own neighbour. Specifically, we choose, $k=6$, which is almost equivalent\footnote{i.e. both methods calculate the density within the same radius, but the number of objects differs by 1, meaning the $k=5$ (not-own-neighbour) number density estimate simply differs by a factor of $5/6$ from the $k=6$ (own-neighbour) estimate.} to the commonly used case where $k=5$ and each galaxy is not considered to be its own neighbour \citep[e.g.][]{Baldry2006}. We test the two methods on a 7~Mpc (proper) slice through the simulation snapshot at $z=0.5$, which corresponds to a difference in velocity due to the Hubble flow of $\Delta$V $=500 $~km s$^{-1}$, and implies a requisite precision in redshift of $\Delta z = 0.002$. Such precision will be achievable at intermediate and high redshift using spectroscopic and grism redshifts from future instruments, such as MOONS \citep{Evans2011}, PFS \citep{Takada2014}, 4MOST \citep{DeJong2012} and JWST \citep{Gardner2006}. We explore estimates of both the 2-D surface density and the 3-D density. We find that the rank of each galaxy indeed remains approximately constant, regardless of either the exact estimator used, or whether we consider the 2-D or 3-D densities. Typically, the rank of a galaxy does not change by more than 10 per cent and, therefore, changing the density estimator leaves our conclusions unchanged. 

\begin{figure*}
  \centering
  \includegraphics[width=1.0\textwidth]{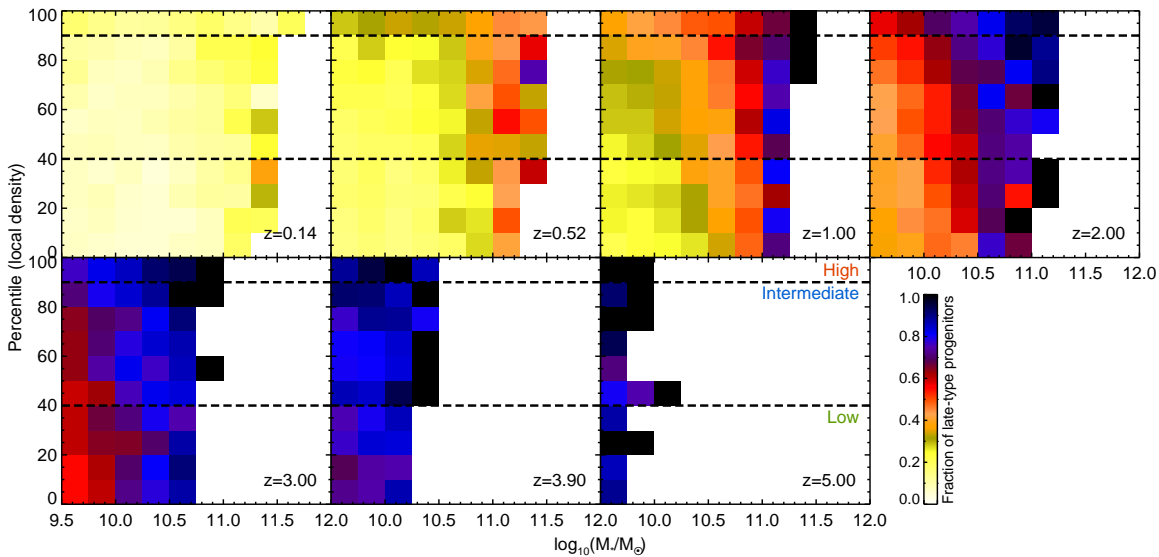}
  \caption{Density plots showing the redshift evolution of the fraction of late-type galaxies that are progenitors of early-types at the present day, as a function of the stellar mass and density percentile of the late-type galaxies in question. The late-type progenitor fraction is represented by the colour bar. We do not plot bins containing 3 or fewer galaxies.}
  \label{progenitorpercentile}
\end{figure*}

Figure \ref{progenitorfraction} shows the fraction of late-type galaxies at a given redshift that are the progenitors of a local early-type, as a function of stellar mass and split by local environment. We show the 2-D progenitor probability as a function of both stellar mass and local density in Figure \ref{progenitorpercentile}, with the colour bar indicating the progenitor probabilities. At all redshifts, there is a positive trend of progenitor probability with stellar mass i.e. more massive late-type galaxies are more likely to be progenitors of local early-type remnants. 

At high redshifts, almost all massive galaxies, regardless of their local environment, are progenitors of present-day early-types. While the progenitor probabilities increase with redshift, for the most massive galaxies the progenitor probability remains close to $\sim1$ until $z\sim0.5$. The principal reason for an increase in the progenitor fraction with redshift is simply the fact that late-type galaxies have more time to merge with other galaxies and undergo morphological transformation before the present day. The rate of morphological transformations is regulated by the merger rate per galaxy, which rises with redshift \citep[e.g.][]{Welker2015,Rodriguez2015,Kaviraj2015a} and thus controls the rate of change in the progenitor fraction as a function of redshift. Lower-mass (M$_\star$ $<$ 10$^{10.5}$~M$_{\odot}$) galaxies can also exhibit high progenitor probabilities at high redshift, but \emph{only} if they occupy regions of high density (e.g. the 80-100$^{\mathrm{th}}$ density percentile, see Figure \ref{progenitorpercentile}). Note that the progenitor probabilities decline for all galaxies towards low redshifts, because these systems will not have had time to undergo enough merging to achieve early-type morphology. 

\begin{figure*}
  \centering
  \includegraphics[width=1.0\textwidth]{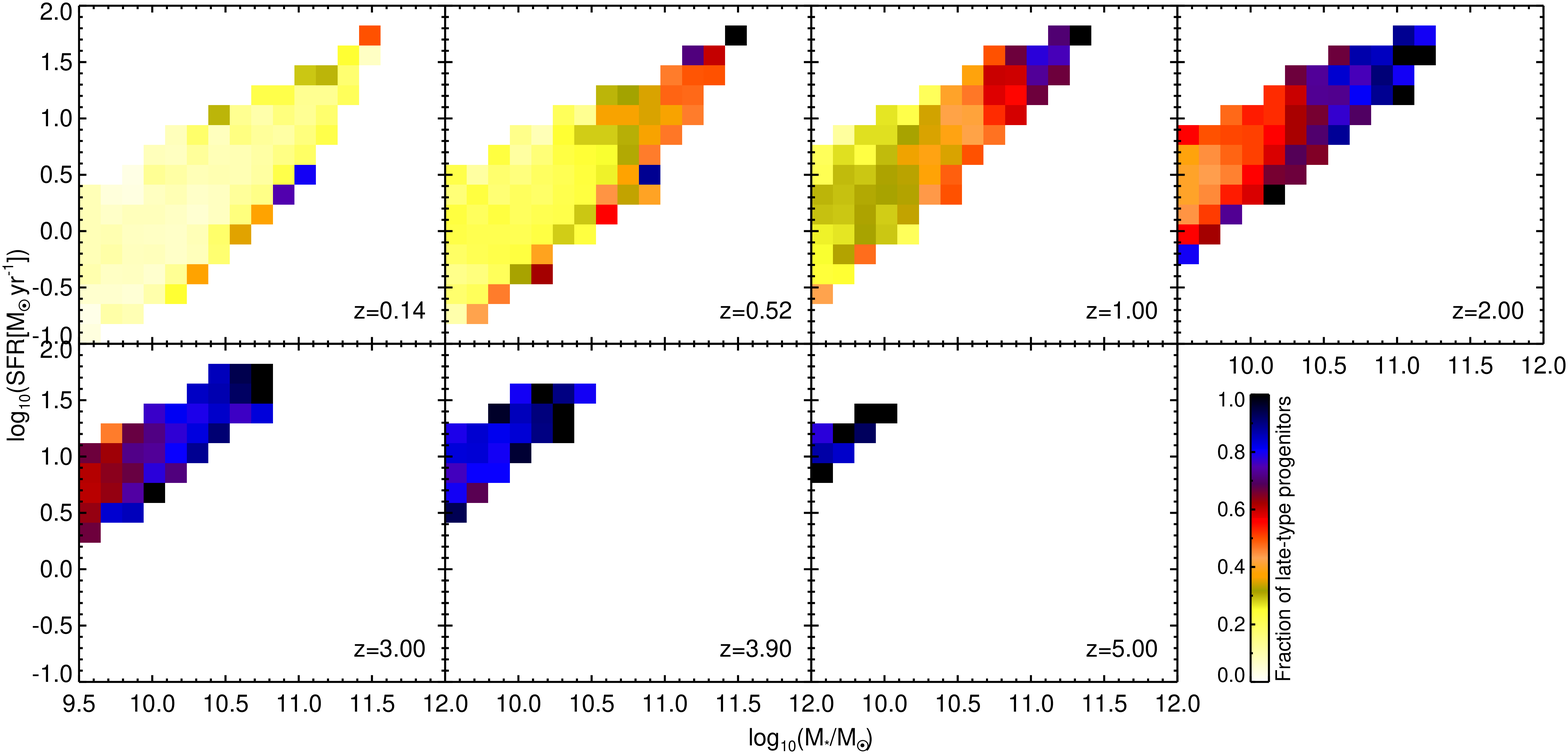}
  \caption{Left: Star-formation main sequence (i.e. star-formation rate plotted against stellar mass) for late-type galaxies, as a function of redshift, colour-coded by the fraction of late-type galaxies that are progenitors of early-types at the present day. We do not plot bins containing 3 or fewer galaxies.}
  \label{progenitorsfr}
\end{figure*}

We proceed, in Figure \ref{progenitorsfr}, by exploring the progenitor probabilities as a function of stellar mass and star-formation rate (the parameter space that is commonly referred to as the `star-formation main sequence'). The colour bar indicates the progenitor probabilities. Mirroring the trends found earlier, massive late-type galaxies are more likely to be progenitors of early-types. At high redshift, these massive late-type progenitors can be some of the most highly star-forming systems in the Universe. However, this is simply a consequence of star-formation activity being, on average, more vigorous in the early Universe. It is worth noting, however, that at all redshifts, progenitor fractions increase \textit{at fixed stellar mass} as the star-formation rate decreases. In other words, at fixed stellar mass, late-type galaxies with lower specific star-formation rates are more likely to be progenitors. For example, at $z\sim2$ (top-right hand panel of Figure \ref{progenitorsfr}), a late-type galaxy with a stellar mass of 10$^{10.7}$~$\mathrm{M}_{\astrosun}$, which resides at the upper end of the star-formation main sequence, has a progenitor probability of around 70 per cent. A galaxy with a similar mass which sits at the bottom of the star-formation main sequence has a progenitor probability of close to 100 per cent. 

We release the progenitor probabilities calculated here, as a function of different observables, as described in the Appendix \ref{sec:appendix}. For a series of redshifts, we provide tables of progenitor probabilities as a function of the three principal observables studied here: stellar mass, star-formation rate and local environment. Since not all observational datasets may offer access to all three quantities, we also provide separate tables for projections of this 3-D parameter space i.e. progenitor probabilities as a function of stellar mass only, stellar mass and local environment and stellar mass and star-formation rate. As noted above, the properties of individual galaxies in current and future observational surveys which provide these observables can be compared to these tables in order to estimate the probability that they are progenitors of an early-type galaxy in the local Universe. The structure of the files containing these tables, and scripts to read them, are provided in the Appendix \ref{sec:appendix}. 

\section{Summary}
\label{sec:Summary}
As end-points of the hierarchical mass assembly process, early-type galaxies host more than half of the stellar mass density in the local Universe, their stellar populations encoding the assembly history of galaxies over cosmic time. Studying these galaxies in the local Universe and probing their progenitors at earlier epochs offers a unique perspective on the evolution of the observable Universe. However, since morphological transformations progressively convert late-type (disc-like) galaxies into early-type systems, the progenitors of today's early-type galaxies become increasingly dominated by late-types at high redshift. Understanding the evolution of early-types over cosmic time therefore requires a reliable method for identifying these late-type progenitors of local early-types. Here, we have used the Horizon-AGN cosmological hydrodynamical simulation, which produces good agreement with the  observed properties of galaxies in the redshift range $0<z<5$, to study how the progenitors of local early-type galaxies evolve over cosmic time.  

We have studied the merger histories of local early-types and the morphologies of galaxies that are involved in these mergers and traced how the morphological mix of galaxies in the progenitor population changes over cosmic time. We have then used the simulation to study the fraction of late-type galaxies that are progenitors of present-day early-types, as a function of redshift, stellar mass, local environment and star-formation rate: observables that can be routinely measured in current and future datasets. As noted earlier, these fractions can be treated as probabilities that can then be used to include late-type progenitors of local early-types in observational surveys by `weighting' these late-type systems by these probabilities. The benefit of this approach is to alleviate progenitor bias i.e. the bias that occurs if one considers only early-type galaxies (or proxies like the red sequence) to study the progenitor population of today's early-types. Our key conclusions are as follows:  

\begin{itemize}

\item The merger history of early-type galaxies indicates that these systems finish assembling their stellar mass at relatively early epochs. By $z\sim1$, around 60 per cent of today's massive early-types, averaged over all environments, have had their last significant merger (i.e. a merger with mass ratio greater than 1:10). For early-type galaxies that inhabit high density environments at the present day (e.g. clusters) this value is 70 per cent, while it is $\sim$50 per cent in early-types that inhabit low-density environments (e.g. the field). On average, morphological transformation is $\sim$50 per cent faster in high-density environments compared to low-density regions.
\\
\item Progenitor bias is significant at all but the lowest redshifts. Until $z\sim0.6$ less than half of the progenitors of today's early-types actually have early-type morphology. Similarly, less than half of the stellar mass that ends up in an early-type today is actually hosted by a progenitor that has early-type morphology at this redshift. Around the epoch of peak cosmic star-formation, which is also the epoch at which morphological transformation occurs most rapidly, studying only early-type galaxies misses almost all (at least 80 per cent) of the stellar mass that eventually ends up in early-types at the present day.  
\\
\item The morphological mix of progenitor galaxies that are involved in mergers evolves over time. At all redshifts, the majority of mergers have at least one late-type progenitor. Mergers between two late-type galaxies dominate at early times i.e. around the epoch of peak cosmic star-formation and beyond ($z>1.5$) and the fraction of mergers involving two early-type galaxies climbs rapidly at low redshift ($z<0.5$). 
\\
\item At all redshifts, late-type galaxies with larger stellar masses are more likely to be progenitors of local early-type remnants. At high redshifts, almost all massive (M$_\star$ $<$ 10$^{11}$~M$_{\odot}$) late-type galaxies, regardless of their local environment, are progenitors of present-day early-type galaxies. While the progenitor probabilities increase with redshift, for these massive galaxies, the progenitor probability remains close to $\sim1$ until $z\sim0.5$. Lower-mass (M$_\star$ $<$ 10$^{10.5}$~M$_{\odot}$) galaxies also exhibit high progenitor probabilities at high redshift, as long as they occupy regions of high density (e.g. the 80-100$^{\mathrm{th}}$ percentiles in density). 
\\
\item At high-redshift, massive late-type galaxies that are progenitors of present-day early-types can be some of the most highly star-forming systems in the Universe, simply because star-formation activity is, on average, more vigorous in the early Universe. However, at fixed stellar mass, progenitor fractions increase as the star-formation rate decreases i.e. late-type galaxies with lower specific star-formation rates are more likely to be progenitors of early-type galaxies.
\end{itemize}

In the impending era of large observational surveys (e.g. LSST, EUCLID, JWST), this paper provides a framework for studies of how the stellar mass hosted by the local early-type galaxy population is built up over cosmic time. 

\section*{Acknowledgements}
We thank the anonymous referee for constructive comments that improved this paper. GM acknowledges support from the Science and Technology Facilities Council [ST/N504105/1]. SK acknowledges a Senior Research Fellowship from Worcester College Oxford. JD acknowledges funding support from Adrian Beecroft, the Oxford Martin School and the STFC. CL is supported by a Beecroft Fellowship. This research has used the DiRAC facility, jointly funded by the STFC and the Large Facilities Capital Fund of BIS, and has been partially supported by grant Spin(e) ANR-13-BS05-0005 of the French ANR. This work was granted access to the HPC resources of CINES under the allocations 2013047012, 2014047012 and 2015047012 made by GENCI. We thank Stephane Rouberol for running and maintaining the Horizon cluster hosted by the Institut d'Astrophysique de Paris. Elias Brinks and Chiaki Kobayashi are thanked for useful comments. This work is part of the Horizon-UK project.




\bibliographystyle{mnras}
\bibliography{references} 




\appendix

\section{Tables of progenitor probabilities}
\label{sec:appendix}

The joint progenitor probability distribution is tabulated as a function of redshift, stellar mass, local environment and star-formation rate and stored in an $a\times b\times c\times d$ binary file. The python routine \texttt{tabulate\_progenitor\_probability.py} outputs progenitor probabilities from the joint probability distribution and can be run as follows:
\\
\\
\noindent \texttt{python tabulate\_progenitor\_probability.py -z <redshift> -m <mass> -p <percentile> -s <SFR>},
\\
\\
\noindent where mass is in units of log$_{10}$(M$_{\star}$/M$_{\astrosun}$) and SFR is in units of M$_{\astrosun}$~yr$^{-1}$. At least 1 of the 4 keywords must be supplied, if fewer than 4 keywords are supplied the routine outputs the progenitor probability marginalised over the missing dimension(s).
\\
\\
e.g. calling `\texttt{python tabulate\_progenitor\_probability.py -z 2.0 -m 10.0 -s 3}' returns the joint progenitor probability for $z=2$, $M_{\star}=10^{10}$~M$_{\odot}$ and $SFR=3$~M$_{\astrosun}$~yr$^{-1}$ with environment marginalised out.\\
\\
\\
We also provide progenitor probabilities in ASCII format. For each redshift $z\in$[0,5] progenitor probabilities are tabulated in 4 files:
\begin{itemize}
\item \texttt{m\_rho\_SFR.txt} contains 4 columns listing stellar mass [log$_{10}$(M$_{\star}$/M$_{\astrosun}$)], percentile, star-formation rate [M$_{\astrosun}$~yr$^{-1}$] and the joint progenitor probability.
\\
\item \texttt{m\_rho.txt} contains 3 columns listing stellar mass, percentile and the progenitor probability marginalised over SFR.
\\
\item \texttt{m\_SFR.txt} contains 3 columns listing stellar mass, SFR and the progenitor probability marginalised over percentile.
\\
\item \texttt{m.txt} contains 2 columns listing stellar mass and the progenitor probability marginalised over percentile and SFR.
\end{itemize}
The routine and files above are available from \url{http://www.star.herts.ac.uk/~gmartin/bias}.


\bsp	
\label{lastpage}
\end{document}